\documentstyle[12pt]{article}

\title{Mode-coupling theory for heteropolymers}
\author{E. Pitard, E.I. Shakhnovich\\
Harvard University\\
Dpt of Chemistry and Chemical Biology\\
12 Oxford Street, MA02138 Cambridge USA}
\date{\today}

\def\be{\begin{equation}}
\def\bea{\begin{eqnarray}}
\def\ee{\end{equation}} 
\def\eea{\end{eqnarray}} 

\begin{document}
\maketitle
\vskip 1cm
\begin{abstract}
We study the Langevin dynamics of a heteropolymer  by means of
a mode-coupling approximation scheme, giving rise to a set of coupled
integro-differential equations relating the response and correlation functions.
The analysis shows that there is a regime at low temperature
characterized by out-of-equilibrium dynamics, with violation
of time-translational invariance and of the fluctuation-dissipation theorem.
The onset of ageing dynamics at low temperatures
gives new insight into the nature of the slow dynamics of a disordered
polymer. We also introduce a renormalization-group treatment of our
mode-coupling equations, which supports our analysis,
and might be applicable to other systems.

\end{abstract}
%\pacs{75.10N, 64.70P, 71.55J}
%{\tt$\backslash$\string pacs\{\}}} 
%%\narrowtext
%\draft
%\bibliographystyle{unsrt}
\vskip 1cm
\noindent\mbox{Submitted for publication to:} \hfill 
\mbox{}\\ \noindent \mbox{``Physical Review E''}\\ 
\vskip 1cm\noindent \mbox{PACS: 83.10.Nn Polymer dynamics; 
83.70.Hq Heterogeneous liquids; 64.70.Pf Glass transition} \newpage

\newpage

%%%%%%%%%%%%%%%%%%%%%%%%%%%
\section{Introduction}
\label{sec: intro}

The dynamics of a heteropolymer chain is relevant for the problem
of protein folding, and also from a fundamental point of view.
Since a protein is composed of monomers of different chemical natures,
it is important to understand the effect of heterogeneity
on the kinetics of a polymer chain. Such results might give insight into
the possible folding pathways of proteins or longer chains.
Although the influence of quenched or annealed disorder
on the thermodynamics of polymers is a largely investigated area of study,
\cite{Garel}
together with the effect of random fields
\cite{EdMuthu}, random charges along the chain
(polyelectrolytes and polyampholytes)
\cite{KardarKantov}, in solution or at the
interface between two fluids
\cite{Interface},
little is known about dynamics.

Previous studies concerning the statics of heteropolymers show
that there exists a frozen phase at low temperature,
 very similar to a spin-glass
phase
\cite{Hetero},
 which is a non-ergodic state characterized by a very slow relaxation.
Recent phenomenological and analytical developments have 
reproduced the experimental evidence of aging in spin-glasses
\cite{revueBou}.
Below a certain temperature, the system relaxes in a slower and slower way
as the waiting time -which is the time elapsed between the beginning of the
experiment and the observation time- is increased; the dependance
on the waiting time is clearly proved experimentally in spin-glasses
and in other glasses \cite{Alba}, \cite{Struik}.
 The relaxation
follows a power law and both time-translational invariance and
the fluctuation dissipation theorem are violated.
Similar properties have been found theoretically
for the study of large-time out of
equilibrium dynamics of a manifold in a random potential
\cite{Doussal}.
These results are of great interest for systems where
disorder or frustration are present, and similar ideas 
start to be applied, for instance for the rheology of soft
 glassy materials \cite{Sollich}, 
or for
the dynamics of structural glasses, where experimental evidence of
the violation of the fluctuation-dissipation theorem has 
recently been reported \cite{Grigera}.

Concerning  the dynamics of heteropolymers, few studies exist
at present \cite {Roan} and some of them show that there may be some
glassy behaviour as the temperature is lowered \cite{THI}, or that
 the relaxation should 
follow a stretched exponential law \cite{Vilgis}.
There is indeed numerical evidence of stretched exponential
relaxation for randomly branched polymers
\cite{Kemp}, or
for the reptation of polymers in disordered media
\cite{Hwa}, as well
as experimental evidence of the same phenomenon
for glasses and proteins \cite{Eaton}.
Moreover, there is a growing literature about the dynamics
of homogeneous but strongly frustrated polymer systems,
such as polymer melts where stretched-exponential laws are observed
through computer simulations
\cite{Baschnagel},
and with dynamics very similar to the one observed in
structural glasses or supercooled liquids \cite{Gotze}.
Other approaches to the problem of 
dynamics of heteropolymers in melts use the concept of reptation
to compute  in a phenomenological way
the relaxation time of a disordered chain
 \cite{BouCates}.

Our work concerns the study of the Langevin dynamics
of a heteropolymer, treated in the Mode-Coupling Approximation scheme (MCA).
Our motivation for using this approximation is based
on several previous studies
that led to significant results.
This procedure goes beyond perturbation theory (though not in 
a very controlled way) and is therefore useful when one wants
to study strong-coupling effects. The scheme is to expand
the microscopic quantities involved in the Langevin equation to
lowest non-trivial order in the potential -as if it was a perturbation
procedure-,
and then to replace in the correction terms the bare correlation functions
(those corresponding to the problem without potential)
by the full correlation functions one wants to compute. This amounts
to resumming a certain class of diagrams and hence to
go further than the weak-coupling regime.
This type of procedure has been used by Kraichnan in the context of turbulence
as a way to find the Kolmogorov laws starting from the Navier-Stokes
equation \cite{Kraichnan};
it has also been used for the KPZ equation where exponents
close to those found by dynamic renormalization group were computed
\cite{KPZ}.
Interestingly, it has also been found
that the MCA is exact for some special models
with quenched disorder which dynamics can be studied
exactly in a mean-field approach, using functional methods;
this is true in particular for the p-spin spherical spin-glass
 model \cite{Physica A}. 
So one can
hope that the MCA approach is able to capture
dynamic effects such as aging that arise from
the presence of disorder in a 
non-perturbative way.
Finally it has been pointed out that the general
coupled set of equations obtained through
MCA looks  very much like those found in the context
of the mode-coupling theory introduced by Gotze
\cite{Gotze}, which
gives a reliable description of the slow dynamics of supercooled liquids,
reinforcing the link between glassy systems which are frustrated but
contain no disorder, and disordered systems such as spin-glasses.

In the following, we show that the same approximation (MCA)
 can be used for the
dynamics
of a disordered polymer and that out-of-equilibrium
features can be found as well. These results may be of relevance for
heteropolymer melts, or for very long chains of heteropolymers.
From a protein-oriented point of view, such results may not be directly
applicable, since they are derived for an infinite 
and purely random system.
However, for large proteins, one may observe some intermediate
slow regime of folding in the globular state, between
the fast initial hydrophobic collapse,
and the final relaxation towards the native state 
once a nucleus \cite{nucleus} has been formed.

\section{The mode-coupling approach}
\label{sec: mode}
\subsection{Formal developments}
\label{sec:formal}

%ALL DETAILED CALCULATIONS IN APPENDICES

We introduce here a model of heteropolymer dynamics
and explain how to derive a set of coupled integro-differential
equations using the Mode Coupling Approximation.
We use a standard hamiltonian for a disordered polymer,
where a quenched potential $V(s, \vec{\phi}(s,t))$
is applied and comes from the random nature 
of the interactions between monomers. 
%We have introduced
%a mass $\mu(t)$ to prevent divergences of correlators if necesssary;
In our notations,
$\vec{\phi}(s,t)$ is the position of monomer $s$ at time $t$, 
$s$ being the coordinate of the monomer along the chain, $s=1,..,N$.
$d$ is the dimensionality of the space and $a_0$ is the Kuhn length.

\begin{equation}
{\cal H}= \frac{1}{2 a_0^2} \int ds \sum_{\alpha=1}^d
\left(\frac{\partial \phi_{\alpha}}{\partial s}\right)^2
%+ \frac{1}{2} \mu(t) \int ds \sum_{\alpha=1}^d \phi_{\alpha}^2(s,t)
+ \int ds V(s, \vec{\phi}(s,t))
\end{equation}

\noindent
More explicitly, the random potential is:

\begin{eqnarray}
&&V(s, \vec{\phi}(s,t))= \int ds' B(s,s')
\delta(\vec{\phi}(s,t)-\vec{\phi}(s',t))\nonumber\\
&&=\int ds' B(s,s') \int \frac{d\vec{q}}{(2\pi)^d}
e^{i \sum_{\alpha} q_{\alpha}[\phi_{\alpha}(s,t)
-\phi_{\alpha}(s',t)]}
\end{eqnarray}

\noindent
We use a bar to perform the average over the disorder,
and we assume that the value of the interaction between too
 different monomers 
is distributed in a gaussian way.

$$\overline{B(s,s')}=0$$
$$\overline{B(s_1,s'_1) B(s_2,s'_2)}= B_0^2 \delta (s_1-s_2)
\delta (s'_1-s'_2)$$

\noindent
We consider the Langevin equation for such a polymer:

\begin{equation}
\left[\frac{\partial}{\partial t} -\frac{1}{a_0^2}\frac{\partial ^2}
{\partial s^2} 
\right]
\phi_{\alpha}(s,t)=
-\frac{\partial}{\partial\phi_{\alpha}(s,t)}
\int ds V(s, \vec{\phi}(s,t))
+\eta_{\alpha}(s,t)
\end{equation}

\noindent
with a gaussian thermal noise  $\vec{\eta}(s,t)$,

$$<\eta_{\alpha}(s,t)>=0$$
$$<\eta_{\alpha}(s,t) \eta_{\beta}(s',t')>= 2T \delta(s-s')
\delta(t-t') \delta_{\alpha \beta}$$

\noindent
Our aim is to compute correlation and response functions,
or at least know their qualitative behaviour with time.
Following earlier studies, we 
don't assume {\it a priori} time-tranlational invariance and
we define respectively the correlation function and the response function
as quantities depending on two distinct times, $t$ and $t'$.

\begin{equation}
C(s,t;s',t')=\frac{1}{d} \sum_{\alpha=1}^{d} \overline{<\phi_{\alpha}(s,t)
\phi_{\alpha}(s',t')>}
\end{equation}

\begin{equation}
G(s,t;s',t')=\frac{1}{d} \sum_{\alpha=1}^{d} \overline{
\frac{\partial <\phi_{\alpha}(s,t)>}{\partial \eta_{\alpha}(s',t')}}
=\frac{1}{d} \frac{1}{2T}\sum_{\alpha=1}^{d} \overline{<\phi_{\alpha}(s,t)
\eta_{\alpha}(s',t')>}
\end{equation}

\noindent
The last identity holds as long as the random noise
$\vec{\eta}(s,t)$ is gaussian.

\noindent
During all this study, we shall use Fourier transforms,
which we define, both for the position $\vec{\phi}(s,t)$
and for the correlation functions, with $\omega_n=\frac{2\pi n}{N}$.

$$\tilde{\phi}_{\alpha}^n (t)=\frac{1}{N} \int_0^N e^{i\omega_n s}
\phi_{\alpha}(s,t)ds$$

$$\tilde{G}^n (t,t')=\frac{1}{N} \int_0^N e^{i\omega_n (s'-s)}
G(s,t;s',t')ds ds'$$

\noindent
The standard procedure in the MCA is to find the solution
 $\tilde{\phi}_{\alpha}^n (t)$ of the Langevin equation,
up to the first non-zero order in perturbation.
The dynamic equation can be rewritten in Fourier space:

\begin{equation}
\frac{\partial \tilde{\phi}_{\alpha}^n (t)}{\partial t}=
-\omega_n^2 \tilde{\phi}_{\alpha}^n (t)
-\lambda \tilde{W}_{\alpha}^n (t) + \tilde{\eta}_{\alpha}^n (t)
\end{equation}

\noindent
where we have added for
convenience the coefficient $\lambda$ as the
perturbative parameter; $\lambda$ is eventually set back to $1$ at the end of
the expansion.
And the quantity $\tilde{W}_{\alpha}^n (t)$ is defined as:

\begin{equation}
\tilde{W}_{\alpha}^n (t)=\frac{2}{N} \int_0^N e^{i \omega_n s} ds
\int_0^N ds' B(s,s') \int \frac{d\vec{q}}{(2\pi)^d}
i q_{\alpha} e^{i \sum_{\alpha} q_{\alpha}
(\phi_{\alpha}(s,t)-\phi_{\alpha}(s',t))}
\end{equation}

\noindent
If $\lambda$ is equal to $0$ we are reduced to the 'bare' problem
of an elastic chain in a harmonic potential.
Then the Langevin equation is exactly solvable
and the solution is

\begin{equation}
\tilde{\phi}_{\alpha,0}^n (t)=\int_0^t dt'
\tilde{G}_0^n (t,t') \tilde{\eta}_{\alpha}^n (t)
\end{equation}

\noindent
where $\tilde{G}_0^n (t,t')$ is the bare response function, and:

\begin{equation}
%\tilde{G}_0^n (t,t')= e^{-\int_{t'}^t d\tau (\mu(\tau)+\omega_n^2)}
\tilde{G}_0^n (t,t')= e^{-\omega_n^2 (t-t')}
\end{equation}

\noindent
When one now adds the disorder-dependant term $\tilde{W}_{\alpha}^n (t)$
in the Langevin equation, then

\begin{equation}
\tilde{\phi}_{\alpha}^n (t)=\int_0^t dt'
\tilde{G}_0^n (t,t') \left[\tilde{\eta}_{\alpha}^n (t')
- \lambda \tilde{W}_{\alpha}^n (t')\right]
\end{equation}

\noindent
is the exact solution, which actually gives an implicit
equation for the quantity $\tilde{\phi}_{\alpha}^n (t)$
that cannot be solved in a straightforward way.
One then performs in this expression an expansion
up to second order in $\lambda$, we refer the reader to Appendix A for
more details.

\noindent
It is then rather straightforward, though computationally
lenghty, to compute  $\tilde{G}^n (t,t')$
and $\tilde{C}^n (t,t')$ as functions of the bare quantities,
which are at the end replaced by the full or
'renormalized' quantities.
One finally ends up with a set of coupled equations which
solutions will in principle fully describe the
dynamics of the original system.
We can write these equations in a compact way:

\begin{eqnarray}
&&\left[\frac{\partial}{\partial t} + \frac{\omega_n^2}{a_0^2} \right]
\tilde{G}^n (t,t')=
\delta(t-t') +
\int_0^t dt_1 R_n (t,t_1) \tilde{G}^n (t,t')\nonumber\\
&&+\int_{t'}^t \Sigma_n (t,t_1) \tilde{G}^n (t_1,t') 
\end{eqnarray}

\begin{eqnarray}
&&\left[\frac{\partial}{\partial t} + \frac{\omega_n^2}{a_0^2} \right]
\tilde{C}^n (t,t')=
\int_0^t dt_1 R_n (t,t_1) \tilde{C}^n (t,t') +
\int_0^{t'} dt_1 D_n(t,t_1) \tilde{G}^n (t',t_1)\nonumber\\
&& +\int_0^t dt_1 \Sigma_n (t,t_1) \tilde{C}^n (t_1,t')
\end{eqnarray}

\noindent
All quantities $R_n (t_1,t_2)$, $\Sigma_n (t_1,t_2)$,
 $D_n(t_1,t_2)$ involved in the coupled set of equations
are defined in Appendix A,  and depend
only on $\tilde{G}^p (t,t')$ and $\tilde{C}^p (t,t')$,
with $p\neq 0$.
Similar sets of coupled equations have been already encounted
for example  in
\cite{CuKu}, \cite{KPZ}, \cite{Doussal}  and can either
be solved numerically or require additional assumptions
to get more information on the solutions.

\subsection{Analysis of the equations}
\label{sec:analysis} 

A very difficult task is to solve the set of integro-differential equations
described above. One major difficulty relies in the fact
that all modes are coupled, and as it has already been observed
for example for
the mode-coupling equations of the KPZ model
\cite{Tu} \cite{KPZ}, the numerical
treatment for these equations presents numerous problems.
We have not been able to make significant progress
in that direction; not only should it require a recursive algorithm
with careful check for the convergence of all functions,
but we also expect some divergences in the long time regime,
which would require the introduction of unknown cut-offs.

\noindent
In our analysis,
we took into account all terms found in the former section,
without truncating them with too crude approximations.
However it would be interesting in the future
to find a way to simplify these equations 
(even if the connection with the initial model
becomes then less obvious),
that would reproduce the results that we find here.

\noindent
The first step in the analysis
can be done by focusing on the large time limit and
proposing an ansatz for  the correlation functions in that time domain.
Let's assume that one can write, in the limit where
$t\rightarrow\infty$ and $t'\rightarrow\infty$, with
$t' \ll t$,

\begin{equation}
\tilde{C}^n (t,t')=q f_n \left(\frac{t'}{t}\right)^{\gamma}
\end{equation}

\begin{equation}
\tilde{G}^n (t,t')=\frac{q'}{t} f_n \left(\frac{t'}{t}\right)^{\gamma-1}
\end{equation}

\noindent
Such an ansatz also contains a generalized version of the fluctuation-dissipation
theorem (this has already
been introduced in earlier studies
\cite{CuKu}), which can be written:

\begin{equation}
\tilde{G}^n (t,t')=x \frac{\partial}{\partial t'}
\tilde{C}^n (t,t')
\end{equation}

\noindent
where $x$ is the coefficient $x=\frac{q'}{\gamma q}$.

\noindent
Then by replacing into the mode-coupling equations, one is left
with implicit equations for $q$, $f_n$, $x$ and $\gamma$, but the dependance
on time cancels out in the limit of large times.
This is what makes the ansatz consistent, at least as far as the dependance on
time is concerned. We give computational
details in Appendix B. 
In particular,
in the equation for  $\tilde{C}^n (t,t')$,
$D_n (t_1,t_2)$ can be written as a sum of four contributions
$D_n^{(i)}(t_1,t_2)$, $i=0,1,2,3$ as shown in Appendix A,
and
the terms involving  $D_n^{(2)}(t_1,t_2)$ and $D_n^{(3)}(t_1,t_2)$
are shown to be negligible in the limit of large times;
this makes the time cancellation possible.
 All parameters $q$, $f_n$, $x$ and $\gamma$ have a dependance on
temperature.
Then the remaining equations 
on $q$, $f_n$, $x$ and $\gamma$
are again too difficult
to solve numerically, since they require the introduction of cut-offs
- see Appendix B -,
to stop the divergences in the integrals when the two time arguments
become too close to each other (in particular, both
$\tilde{G}^n (t,t')$ and 
$\tilde{C}^n (t,t')$ have different analytical forms when
$t'\rightarrow t$, which we don't know).

\medskip

\noindent
The two extreme cases $T=0$ and $T=\infty$ are of interest
in this problem.
In the case where the temperature is zero,
the equations can be simplified,
the terms involving $D_n^{(2)}(t_1,t_2)$ and $D_n^{(3)}(t_1,t_2)$
are equal to zero and one sees easily that
 the above ansatz still remains valid, for the same reasons as the ones
explained above (see Appendix B).

If we study the limit $T\rightarrow\infty$,
one is left with a single term in the equation for $\tilde{C}^n (t,t')$:

\begin{equation}
\int_0^{t'} dt_1 D_n^{(3)}(t,t_1) \tilde{G}^n (t',t_1)=0
\end{equation}

\noindent
The power-law ansatz used above can no longer
satisfy this condition.

\noindent
However, if one assumes time-tranlational invariance and the usual
fluctuation-dissipation theorem 
($  G(\tau)=-\frac{\theta(\tau)}{T} \partial_{\tau} C(\tau)$)
%then one can neglect the term involving  $D_n(t_1,t_2)$
the mode-coupling equations are simply satisfied.
In the limit of infinite temperature, one is actually left
with the simple Rouse model for a homopolymer chain, and an exponential time
relaxation.

Such information about  a glassy behaviour at low temperatures
with power-law behaviour lead naturally to think of
the existence of a critical temperature $T_c$ that may separate
the glassy, non-FDT, non-TTI regime from a high temperature
regime where the relaxation would be typically exponential.
Although it is not possible to determine this temperature
from our equations, it should be easier to observe
such a phenomenon in simpler models of polymers,
or by studying numerical models of polymers.

\section{An alternative treatment of the mode-coupling equations:
Functional renormalization group approach}
\label{sec: rgroup}

In view of the difficulties raised by
the mode-coupling equations, one has to search for new analytical
methods to try and solve them. One of them is to
apply a functional renormalization group analysis
to the mode-coupling equations themselves.
To our knowledge, such a method has never been used in this context.
For the present problem, this procedure enables us to have more
information about the analytical form of the correlation functions.
In particular it can justify some scaling form for their
analytical expressions, as soon as one finds a fixed point
in the RG procedure that is believed to represent the small frequency,
small wavevector regime. This is motivated by the fact that we are mostly
 interested
in the long times limit, and in the long-distance regime
($s-s'\rightarrow\infty$, along the chain).

In the case of the disordered polymer, our 
dynamic RG calculation gives rise to a fixed
point.
However, the fixed points equations are themselves hard to solve.
We also believe that such a method could be of interest for
simpler and largely studied systems, such as the KPZ equation.

In this calculation
\cite{HoHal}, we want to integrate out the fast wavevector modes
and keep only the slow modes, the ones with small wavevectors. One also
has to do the same  for high frequencies, to keep track only of the low frequency
part; this can be done by expressing all the quantities in frequency space, but we
didn't describe it
here for simplicity.
For that purpose, we first switch from discrete to continuous Fourier
variables, by replacing $\sum_{n}$ by $\int \frac{d^d k}{2 \pi}$
with an upper cut-off $\Lambda$. The wave vectors such that
$\frac{\Lambda}{b}<||\vec k||< \Lambda$ can be integrated out, with 
$b=e^{\delta l}$ being close to 1; then the only perturbative parameter here is
$\delta l$. After integration, we denote the quantities  $Q$
for which
the $||\vec{k}||$ integration is only
$\int_0^{\Lambda/b} \frac{d^d k}{2 \pi}$ by $Q^{<}$.

In a more convenient way than the ones used in the previous section, 
the starting mode-coupling equations we used can be rewritten 
in the form \cite {Physica A} (see also Appendix A):

\begin{equation}
C_k(t,t')=C^0_k(t,t')+\int_0^t dt_1 \int_0^{t'} dt_2
G_k(t,t_1)D_k(t_1,t_2)G_k(t',t_2)
\end{equation}

\begin{equation}
G_k(t,t')=G^0_k(t,t')+\int_0^t dt_1 G^0_k(t,t_1)
\left[ \int_0^{t_1} dt_2 R_k(t_1,t_2)G_k(t_1,t')
 +\int_0^{t_1} dt_2  \Sigma_k(t_1,t_2)G_k(t_2,t')\right]
\end{equation}

In such a way, we can write for a quantity $Q_k(t_1,t_2)$, the following one-order
expansion
$(Q= G, C, R, \Sigma, D)$:

\begin{equation}
Q_k(t_1,t_2)=Q_k^{<}(t_1,t_2) -\delta l {\cal Q}_k(t_1,t_2)
\end{equation}

The corrections ${\cal Q}_k(t_1,t_2)$ can be computed and only involve
the different quantities $G_0$ and $Q^{<}$.
In particular, for ${\cal G}_k(t,t')$ and ${\cal C}_k(t,t')$, we have:

\begin{eqnarray}
&&{\cal G}_k(t,t')=\int_0^t dt_1 G^0_k(t,t_1)
[ \int_0^{t_1} dt_2 {\cal R}_k(t_1,t_2)G_k(t_1,t')
 +\int_0^{t_1} dt_2  {\cal S}_k(t_1,t_2)G_k(t_2,t') \nonumber \\
&&+\int_0^{t_1} dt_2  R_k(t_1,t_2){\cal G}_k(t_1,t')
 +\int_0^{t_1} dt_2  \Sigma_k(t_1,t_2){\cal G}_k(t_2,t')
]
\end{eqnarray}

\begin{eqnarray}
&&{\cal C}_k(t,t')=\int_0^t dt_1 \int_0^{t'} dt_2
[G_k(t,t_1)D_k(t_1,t_2){\cal G}_k(t',t_2)\nonumber\\
&&+G_k(t,t_1){\cal D}_k(t_1,t_2)G_k(t',t_2)
+{\cal G}_k(t,t_1)D_k(t_1,t_2)G_k(t',t_2)
]
\end{eqnarray}

%DETAILS FOR THE OTHERS?APPENDIX C

%\noindent
%We give more details and the corrections to the
%other quantities in Appendix C.

\noindent
The next step is to do a rescaling of all quantities and write a differential
equation
for the renormalization flow where the increment is $\delta l$.
One has then to make scaling assumptions, which are expected, at least at the
fixed points,
if any. Then, if one assumes

\begin{equation}
Q_k(t,t')=\frac{1}{k^{\chi}} q(t k^z, t' k^z)
\end{equation}

\noindent
where $z$ is the dynamic exponent,
the renormalized quantity is

\begin{equation}
Q_k^R (t,t')=b^{\chi} Q_{bk}^{<}(b^{-z}t,b^{-z}t')
\end{equation}

\noindent
By expanding this last expression to first order in $\delta l$ one finally ends up
with:

\begin{equation}
\frac{\partial Q_k}{\partial l}= \chi Q_k (t,t') + k \frac{\partial Q_k}{\partial k}
-zt \frac{\partial Q_k}{\partial t} -zt' \frac{\partial Q_k}{\partial t'}
+{\cal Q}_k(t,t')
\end{equation}

\noindent
More specifically we used 
for the $q$ functions the scaling forms that correspond to the ansatz of the
preceding section (see equations (13) and (14)),
written in such a way that the values of $\chi$
can be simply identified,

\begin{equation}
C_k(t,t')=\frac{1}{k^{2\alpha +d -z}} c \left(\frac{t'}{t}\right)
\end{equation}

\begin{equation}
G_k(t,t')=\frac{1}{k^{-z}}  \frac{1}{tk^z} g \left(\frac{t'}{t}\right)
\end{equation}

\begin{equation}
D_k(t,t')=\frac{1}{k^{2\alpha +d -3z}} d \left(\frac{t'}{t}\right)
\end{equation}

\begin{equation}
R_k(t,t')=\frac{1}{k^{-z}}  \frac{1}{tk^z} r \left(\frac{t'}{t}\right)
\end{equation}

\begin{equation}
\Sigma_k(t,t')=\frac{1}{k^{-z}}  \frac{1}{tk^z} s \left(\frac{t'}{t}\right)
\end{equation}

Due to the scaling nature
of all quantities,
the derivative terms 
$\frac{\partial Q_k}{\partial k}$ and 
$\frac{\partial Q_k}{\partial t}$ in the flow equations can be simplified 
and expressed in terms of  $Q_k (t,t')$, and we end up with
the following set:

\begin{equation}
\frac{\partial G_k}{\partial l}= z G_k (t,t')+{\cal G}_k(t,t')
\end{equation}

\begin{equation}
\frac{\partial C_k}{\partial l}= -2(d+2\alpha -z)  C_k (t,t')+{\cal C}_k(t,t')
\end{equation}

\begin{equation}
\frac{\partial D_k}{\partial l}= -2(d+2\alpha -3z) D_k (t,t')+{\cal D}_k(t,t')
\end{equation}

\begin{equation}
\frac{\partial R_k}{\partial l}= z R_k (t,t')+{\cal R}_k(t,t')
\end{equation}

\begin{equation}
\frac{\partial \Sigma_k}{\partial l}= z \Sigma_k (t,t')+{\cal S}_k(t,t')
\end{equation}

The fixed points are obtained by setting the $l$ derivative to 0,
and denoting by $Q^{*}$
the fixed-point quantities,
and replacing in the expressions for
 ${\cal G}_k(t,t')$ and ${\cal C}_k(t,t')$,
we
obtain the self-consistent equations at the fixed point.

\begin{equation}
G_k^{*}(t,t')=\int_0^t dt_1 G^0_k(t,t_1)
\left[ \int_0^{t_1} dt_2 R^{*}_k(t_1,t_2)G^{*}_k(t_1,t')
 +\int_0^{t_1} dt_2  \Sigma^{*}_k(t_1,t_2)G^{*}_k(t_2,t')\right]
\end{equation}

\begin{equation}
(d+2\alpha -z) C_k^{*}(t,t')= (d+2\alpha -4z) \int_0^t dt_1 \int_0^{t'} dt_2
G_k^{*}(t,t_1)D_k^{*}(t_1,t_2)G_k^{*}(t',t_2)
\end{equation}

We  checked that the power-law ansatz used
in the first section is still a solution for these 
equations of fixed point;
the procedure of replacing the ansatz
in the expressions is exactly the same as the
one described in Appendix B.
This can then justify its use
in the  mode-coupling equations (11) and (12) 
from which we started our analysis, and which represent the real-space quantites
of interest; the scaling form of the ansatz is also
justified.
If we then
plug at the fixed point
the generalized fluctuation-dissipation theorem
$\tilde{G}^n (t,t')=x \frac{\partial}{\partial t'}
\tilde{C}^n (t,t')$,
we obtain the following relation between exponents:
$2\alpha + d= 2z$.
Further information on the exponents
$\alpha$ and $z$ could be obtained 
by solving numerically coupled equations like (35) and (36);
this will be done elsewhere.

\section{Discussion and Conclusion}
\label{sec:concl}

The analysis of the mode-coupling equations using
renormalization group ideas
enabled us to write the dynamic correlation functions
of a disordered polymer in terms of scaling functions.
Unfortunately, at this point, it does not give much insight
into the exact asymptotic time and wavevector dependance, at large times
and large wavelenghts. We believe however that this
new method could be very convenient for more
simple examples of stochastic equations, that are treated
using mode-coupling techniques.

\noindent
The main result of the mode-coupling approach
 is the evidence of the
out-of-equilibrium character of the dynamics of heteropolymers,
in the thermodynamic limit. The dependence of
 the dynamic correlation functions on
two times -the smaller one being the waiting time-,
implying no time-translational invariance and
a generalized fluctuation-dissipation theorem, is required at low temperatures;
more precisely, a power-law dependence on two times
is a solution of our set of equations.
This contains the aging phenomenon, similar to the one
observed in spin-glasses and other types of disordered systems.

\noindent
Similar calculations as the one described above can be made,
 by slightly changing 
the starting Langevin equation.
We replaced the disordered interaction $B(s,s')$ between monomers
by a constant attractive interaction coefficient $v$, therefore restricting
our study to homopolymers. Remarkably enough, we found that the
power-law ansatz still holds at low temperature. Even in
 the absence of disorder,
the mode-coupling
equations seem to induce some apparent disorder in frustrated 
systems (such as polymer chains in the collapsed phase in this case),
and lead to non-trivial glassy behaviour.
In the MCA approach, it appears that one can capture
the frustration character of a collapsed
polymeric chain that stems from the competition between the attractive
interaction between monomers
and the harmonic potential due to the elasticity of the chain 
(the classical Mode Coupling Theory \cite{Gotze}
for glasses also captures frustration in such simple realizations
as binary  mixtures of Lennard-Jones -LJ- particles).

%COMMENT ON BASCHNAGEL:
%POLYMER/FRUSTRATED LJ/ MICROSCOPIC
%ORIGIN OF GLASS/G1(t) ET FACTEUR DE STRUCTURE

%DISORDERED POLYMER/HOMOPOLYMER

%HOMOPOLYMER: COMPARER DOI-EDWARDS ET MCA
%MCA->STRONG COUPLING, STRONG DENSITY, MELT
%ROUSE MODIFIE, DOI-EDWARDS, 1 CHAINE, EXPONENTIEL

%HOMOPOLYMER/FRUSTRATED PARTICLES IN RANDOM POTENTIAL
%SLOW FOR HIGH DENSITIES (MCA)

%FRUSTRATED PARTICLES/ SAME COLLAPSING PARTICLES:
%GLASS/MODELE TRIVIAL QU'IL EST SANS INTERET DE TRAITER PAR MCA.

%DESCRIPTIoN PLUS MICROSCOPIQUE

\noindent
This similarity of behaviour
between disordered and non-disordered polymers
is not very surprising when one remembers that
mode-coupling equations have very similar
structures in the disordered and non-disordered version of
a given model \cite {Physica A}.
More precisely, it has been previously shown
that given a non-disordered but frustrated or chaotic model, one can find a
disordered model of the same class whose mode-coupling
equations are actually exact, and coincide with
the approximate mode-coupling equations of the non-disordered model.
(This is however not exactly the situation that occurs
here, since the mode-coupling closure is only
an approximation scheme for our heteropolymer model.)
This emphasizes the importance of such studies for
real glasses, which are not characterized by the presence of disorder.
In particular, our analysis may be helpful to understand
the dynamics of polymeric glasses, or melts, that also
exhibit aging \cite {Struik} and vitreous
dynamics \cite{Baschnagel}, qualitatively very similar
to the physics of binary Lennard-Jones fluid mixtures
\cite{Kob}.

\noindent
It is interesting to note that a great similarity has been
observed in the dynamics of supercooled liquids (
modelled by binary mixtures
LJ particles) and of polymer melts,
whereas a few differences are noticeable due to 
the connectivity effect of the polymeric system \cite{Baschnagel}.
It has been shown that the correlation functions
and in particular the dynamic structure factor 
for polymer melts can be well described
by the MCT for liquids. However the mean-square
deplacement shows a subdiffusive behaviour compared to the
case of non-connected LJ particles, due to the connectivity of the
chains. This difference occurs in the 
$\alpha$ regime, at large times, that reflects the rearrangements of the chain
at scales larger that the size of a particule's cage,
whereas the $\beta$ regime, at small times, corresponds to local movements
inside a cage, which
are the same in the presence or not of connectivity.
On the ground of these microscopic
mechanisms, it does not seem surprising either,
that a disordered
polymer melt or a homogeneous melt may have the same kind of dynamics.

\noindent
To summarize, our model provides a tentative theoretical framework
which allows to exhibit slow dynamics and aging
in dense phases of homogeneous or disordered polymers.
A further step in this study would be to
find an efficient numerical algorithm for
the resolution of the mode-coupling equations
or of some simplified form of these mode-coupling equations.
This would provide the full solutions for
the correlation functions, as well as for the mean-square displacement,
as functions of time, and allow for a direct 
comparison with the numerical results of Baschnagel and al.
\cite{Baschnagel}.

%QUANTITATIVE COMPARISON WITH BASCHNAGEL:
%EFFICIENT ALGORITHM? G1(t)?

\noindent
An important issue that also
needs to be clarified is in  which way
the fluctuation-dissipation theorem is violated at long times.
It has been shown very recently
\cite{Peliti} that in the case of systems
with short range interactions, the measurement
of the violation of the dissipation-fluctuation ratio is directly
related to the pattern of the
replica symmetry breaking (`one-step rsb' or `full rsb').
The statics of the heteropolymer model
studied here were reported in
\cite{Hetero} and showed a one-step replica symmetry breaking, so
the fluctuation-dissipation should,
according to \cite{Peliti} be of the form given by equation (15).
In addition, concerning the case of homopolymers,
it would be of great interest
to use the first principle computations
described in \cite {Mezard} and be able to draw a link
between the equilibrium and the dynamical properties in the glass phase.

%KOB/JL BARRAT: AGING +RELATION WITH RSB?

%EXPERIMENTAL REALIZATION OF LJ PARTICLES:
%WEEKS/ S. GLOTZER.

%\begin{itemize}
%\item({i}) 
%\item({ii}) 
%\end{itemize}

\bigskip

E.P. wishes to thank J-P Bouchaud, L. Cugliandolo and J. Kurchan
for interesting and stimulating discussions.
E.P. was supported during this work
by NIH grant GM 52126.

\newpage

{\bf Appendix A}

%LABELS!

In this appendix, we give the full expressions for the 
mode-coupling equations obtained in section 2, as well as some intermediate
results.

Starting from equation (10),
one performs an expansion to second order in $\lambda$,
which leads to an approximate expression for 
$\tilde{\phi}_{\alpha}^n (t)$.
This expression is the following:

\begin{eqnarray}
&&\tilde{\phi}_{\alpha}^n (t)=\tilde{\phi}_{\alpha,0}^n (t)  
-\frac{2\lambda}{N} \int_0^t dt' \tilde{G}_0^n (t,t')
W_{\alpha,0}^n(t')
\nonumber\\
&&+\frac{2\lambda^2}{N} \int_0^t dt' \tilde{G}_0^n (t,t')
\int_0^N ds \int_0^N ds' e^{i\omega_n s} B(s,s')\nonumber\\
&&\int \frac{d^d q}{(2\pi)^d} i q_{\alpha}
e^{i\sum_{\alpha}q_{\alpha} \sum_{m\neq 0}
(e^{-i\omega_m s}-e^{-i\omega_m s'})\tilde{\phi}_{\alpha,0}^m (t')}
\nonumber\\
&&\sum_{\beta} i q_{\beta} \sum_{p\neq 0}
(e^{-i\omega_p s}-e^{-i\omega_p s'})
\int_0^{t'} dt'' \tilde{G}_0^p (t',t'')
W_{\beta,0}^p(t'')\nonumber
\end{eqnarray}

\noindent
where
$$W_{\alpha,0}^n(t')=
\int_0^N ds \int_0^N ds' e^{i\omega_n s} B(s,s')
\int \frac{d^d q}{(2\pi)^d} i q_{\alpha}
e^{i\sum_{\alpha}q_{\alpha} \sum_{m\neq 0}
(e^{-i\omega_m s}-e^{-i\omega_m s'})\tilde{\phi}_{\alpha,0}^m (t')}$$

\noindent
Starting from this expression we calculated 
$\tilde{G}^n (t,t')$ and
$\tilde{C}^n (t,t')$, performing both averages
on the random noise and on the disorder.
The result can be written in terms of
$\tilde{G}_0^p (t,t')$
and $\tilde{C}_0^p (t,t')$ only; but the MCA
approximation consists in replacing these
quantities by the full unknown quantities
$\tilde{G}^p (t,t')$
and $\tilde{C}^p (t,t')$,
resulting in a set of coupled Dyson equations:

$$
\tilde{C}_n(t,t')=\tilde{C}^0_n(t,t')+\int_0^t dt_1 \int_0^{t'} dt_2
\tilde{G}_n(t,t_1)D_n(t_1,t_2)\tilde{G}_n(t',t_2)
$$

$$
\tilde{G}_n(t,t')=\tilde{G}^0_n(t,t')+\int_0^t dt_1 \tilde{G}^0_n(t,t_1)
\left[ \int_0^{t_1} dt_2 R_n(t_1,t_2)\tilde{G}_n(t_1,t')
 +\int_0^{t_1} dt_2  \Sigma_n(t_1,t_2)\tilde{G}_n(t_2,t')\right]
$$

\noindent
The mode-coupling equations
can be rewritten as integro-differential
equations, performing on the left and right-hand sides of the 
previous equations
a convolution with
${\tilde{G_0}}^{-1}$.
The resulting set of equations is the one introduced in section 2:

\begin{eqnarray}
&&\left[\frac{\partial}{\partial t} +\frac{\omega_n^2}{a_0^2} \right]
\tilde{G}^n (t,t')=
\delta(t-t') +
\int_0^t dt_1 R_n (t,t_1) \tilde{G}^n (t,t')\nonumber\\
&&+\int_{t'}^t \Sigma_n (t,t_1) \tilde{G}^n (t_1,t')\nonumber 
\end{eqnarray}

\begin{eqnarray}
&&\left[\frac{\partial}{\partial t} +\frac{\omega_n^2}{a_0^2} \right]
\tilde{C}^n (t,t')=
\int_0^t dt_1 R_n (t,t_1) \tilde{C}^n (t,t') +
\int_0^{t'} dt_1 D_n(t,t_1) \tilde{G}^n (t',t_1)\nonumber\\
&& +\int_0^t dt_1 \Sigma_n (t,t_1) \tilde{C}^n (t_1,t')\nonumber
\end{eqnarray}

\noindent
The term $\int_0^t dt_1 R_n (t,t_1)$ plays the role of a mass and is given by
the following expression:

\begin{eqnarray}
&&R_n(t_1,t_2)=-\frac{2 B_0^2}{N^3} \frac{1}{(2\pi)^d}
\int_0^N ds_1 \int_0^N ds'_1 (1-e^{i\omega_n (s'_1-s_1)})
\sum_{p \neq 0}(1-e^{i\omega_p (s_1-s'_1)})
\tilde{G}^p (t_1,t_2)\nonumber\\
&&\left[\Delta(s_1,s'_1;t_1,t_2)\right]^{-2-\frac{d}{2}}
3 X(s_1,s'_1;t_1) Y(s_1,s'_1;t_1,t_2)\nonumber
\end{eqnarray}

\noindent
The quantity $\Sigma_n(t_1,t_2)$ can be seen as the self-energy associated
with the corresponding Dyson equations:

\begin{eqnarray}
&&\Sigma_n(t_1,t_2)=\frac{2 B_0^2}{N^3} \frac{1}{(2\pi)^d}
\int_0^N ds_1 \int_0^N ds'_1 (1-e^{i\omega_n (s'_1-s_1)})
\sum_{p \neq 0}(1-e^{i\omega_p (s_1-s'_1)})
\tilde{G}^p (t_1,t_2)\nonumber\\
&&\left[\Delta(s_1,s'_1;t_1,t_2)\right]^{-2-\frac{d}{2}}
\left[ X(s_1,s'_1;t_1) X(s_1,s'_1;t_2) 
+ 2 Y^2 (s_1,s'_1;t_1,t_2)\right]\nonumber
\end{eqnarray}

\noindent
Finally, $D_n(t_1,t_2)$ can be seen as a 'renormalized' 
noise correlator, which expression is reported here:

$$
D_n(t_1,t_2)=D_n^{(0)}(t_1,t_2)+
D_n^{(1)}(t_1,t_2)+
D_n^{(2)}(t_1,t_2)+
D_n^{(3)}(t_1,t_2)
$$

\noindent
Each term can be computed separately:

$$
D_n^{(0)}(t_1,t_2)=\frac{2T}{N} \delta(t_1-t_2)
$$

\begin{eqnarray}
&&D_n^{(1)}(t_1,t_2)=\frac{4B_0^2}{N^2} 
\frac{1}{(2\pi)^d}
\int_0^N ds_1 \int_0^N ds'_1
Y(s_1,s'_1;t_1,t_2)\nonumber\\
&&\left[\Delta(s_1,s'_1;t_1,t_2)\right]^{-1-\frac{d}{2}}\nonumber
\end{eqnarray}

\begin{eqnarray}
&&D_n^{(2)}(t_1,t_2)=\frac{2B_0^2}{N^2} 
\frac{2T}{N}
\frac{1}{(2\pi)^d}
\int_0^N ds_1 \int_0^N ds'_1 (1-e^{i\omega_n (s'_1-s_1)})\nonumber\\
&&\sum_{p \neq 0} \int_0^{t_2} dt_3
\tilde{G}^p (t_2,t_3) (1-e^{i\omega_n (s_1-s'_1)})
\left[\Delta(s_1,s'_1;t_1,t_2)\right]^{-2-\frac{d}{2}} \nonumber\\
&&( -3 X(s_1,s'_1;t_3) Y(s_1,s'_1;t_2,t_3)
\int_0^{t_2} dt_4 \delta (t_1-t_4) \tilde{G}^n (t_2,t_4)\nonumber\\
&&+\left[X(s_1,s'_1;t_2)X(s_1,s'_1;t_3) +2Y^2(s_1,s'_1;t_2,t_3)\right]
\int_0^{t_3} dt_4 \delta (t_1-t_4) \tilde{G}^n (t_3,t_4) )\nonumber
\end{eqnarray}

$$
D_n^{(3)} (t_1,t_2)=D_n^{(2)*} (t_2,t_1)
$$

\noindent
Finally we used the following notations:

$$
X(s_1,s'_1;t_1)=\sum_{m \neq 0} |e^{-i\omega_m s_1}
- e^{-i\omega_m s'_1}|^2 \tilde{C}^m (t_1,t_1)
$$

$$
Y(s_1,s'_1;t_1,t_2)=\sum_{m \neq 0} |e^{-i\omega_m s_1}
- e^{-i\omega_m s'_1}|^2 \tilde{C}^m (t_1,t_2)
$$

$$
\Delta(s_1,s'_1;t_1,t_2)=
X(s_1,s'_1;t_1) X(s_1,s'_1;t_2)-
Y^2 (s_1,s'_1;t_1,t_2)
$$

\newpage

{\bf Appendix B}

In this appendix, we show how the ansatz proposed
by equations (13) and (14) can be a suitable solution for
the mode-coupling equations obtained in section 2.

We give the expressions of the different terms introduced in Appendix A,
after performing adequate change of variables, after
the power-law ansatz has been pluged into these terms.
Whenever required, we introduced lower and upper cut-offs,
assuming that in the extreme regimes of times (short times
or $t'\rightarrow t$, for which
we don't have explicit analytical forms)
 the integrals are actually convergent, and that the time
scaling we find is not modified by these contributions.
In all equations we neglect the derivative terms, which is justified
in the limit $t' \ll t$.

$1 \bullet$ Equation for $\tilde{G}_n(t,t')$

\begin{itemize}
\item({i}) 
The first term in front of the propagator
in the right-hand side of the equation is comparable to a mass:
\begin{eqnarray}
&&\int_0^t dt_1 R_n (t,t_1)=-\frac{2B_0^2}{N^2} \frac{1}{(2\pi)^d}
\int_0^N ds (1-e^{-i\omega_n s})
\frac{3q'}{2q} [X(s)]^{-1-d}\nonumber\\
&&\int_0^{1-\epsilon} du u^{-1+2\gamma} (1-u^{2 \gamma})^{-2-\frac{d}{2}}
\nonumber
\end{eqnarray}

This
factor is independant of time, up to the
cut-off $1-\epsilon$.
\item({ii}) 
Concerning the second term,
it is in fact proportional to $\tilde{G}^n (t,t')$:
\begin{eqnarray}
&&\int_{t'}^t \Sigma_n (t,t_1) \tilde{G}^n (t_1,t')=
\frac{2B_0^2}{N^2} \frac{1}{(2\pi)^d}
\int_0^N ds (1-e^{-i\omega_n s})
\frac{q'^2 f_n}{2q} [X(s)]^{-1-d}\nonumber\\
&&\int_{\frac{t'}{t}>w}^{1-\epsilon} du u^{-1} (1-u^{2 \gamma})^{-2-\frac{d}{2}}
[1+u^{2\gamma}] \frac{1}{t} 
\left(\frac{t'}{t}\right)^{\gamma -1}\nonumber
\end{eqnarray}
Finally, given the cut-offs $w$ and $1-\epsilon$,
 all terms in the equation  for $\tilde{G}^n (t,t')$ are
proportional to each other,
and hence the dependance on time cancels out.
\end{itemize}

$2 \bullet$ Equation for $\tilde{C}_n(t,t')$

\begin{itemize}
\item({i})
The first new term that has to be computed is  in fact proportional to
 $\tilde{C}^n (t,t')$:
\begin{eqnarray}
&&\int_{t'}^t \Sigma_n (t,t_1) \tilde{C}^n (t_1,t')=
\frac{2B_0^2}{N^2} \frac{1}{(2\pi)^d}
\int_0^N ds (1-e^{-i\omega_n s})
\frac{q'f_n}{2} [X(s)]^{-1-d}\nonumber\\
&&\int_{w}^{1-\epsilon} du u^{-1} (1-u^{2 \gamma})^{-2-\frac{d}{2}}
[1+2u^{2\gamma}] 
\left(\frac{t'}{t}\right)^{\gamma}\nonumber
\end{eqnarray}
\item({ii}) 
The remaining terms coming from the 
$D_n (t,t_1)$ contribution can be simplified. First, when
$t'<t$, it is easy to see that the ones involving $D_n^{(0)} (t,t_1)$
and $D_n^{(2)} (t,t_1)$ are equally zero.
\item({iii})
The $D_n^{(1)} (t,t_1)$ contribution is again proportional
to $\tilde{C}^n (t,t')$,
$$\int_0^{t'} dt_1 D_n^{(1)}(t,t_1) \tilde{G}^n (t',t_1)=
\frac{4B_0^2}{N} \frac{1}{(2\pi)^d}
\int_0^N ds 
q'f_n [X(s)]^{-1-d}
\frac{1}{2\gamma} \left(\frac{t'}{t}\right)^{\gamma}$$
\item({iv})
Finally, the last term has a different time dependance.
\begin{eqnarray}
&&\int_0^{t'} dt_1 D_n^{(3)}(t,t_1) \tilde{G}^n (t',t_1)=
\frac{2B_0^2 T}{N^2} \frac{1}{(2\pi)^d}
 \int_0^N ds (1-e^{-i\omega_n s})
\frac{q'}{q}(q'f_n)^2 [X(s)]^{-1-d}\nonumber\\
&&\left(-3\int_w^1 du u^{2\gamma -2}
\int _0^{1-\epsilon} dv v^{2\gamma -1} [1-v^{2\gamma}]^{-2-\frac{d}{2}}
+ \int_w^1 du u^{2\gamma -2}
\int _w^{1-\epsilon} dv v^{-1} [1-v^{2\gamma}]^{-2-\frac{d}{2}}
\right)\nonumber\\
&& \frac{1}{t} \left(\frac{t'}{t}\right)^{\gamma-1}\nonumber
\end{eqnarray}
Because of its scaling behaviour in $t$ and $t'$,
this term is in fact a sub-dominant term
compared to $\tilde{C}^n (t,t')$
in the limit $t\rightarrow\infty$ and $t'\rightarrow\infty$, 
$t' \ll t$, and we can neglect it.
\end{itemize}

\newpage

%{\bf Appendix C}

%We present in this appendix a more complete set of results
%and intermediate computations for the RG approach to the mode-coupling
%equations presented in section 3.

\newpage

%\begin{references}

\bibliographystyle{unsrt}

%\end{references}

\end{document}